\documentclass[12pt]{article}
\usepackage{epsfig}
\usepackage{latexsym}
\DeclareGraphicsExtensions{.eps.gz,.eps,.ps,.ps.gz}
\parskip=\itemsep               %?
\date{}
\catcode`@=11
\def\@citex[#1]#2{\if@filesw\immediate\write\@auxout{\string\citation{#2}}\fi
  \def\@citea{}\@cite{\@for\@citeb:=#2\do
    {\@citea\def\@citea{,\penalty\@m}\@ifundefined
      {b@\@citeb}{{\bf ?}\@warning
       {Citation `\@citeb' on page \thepage \space undefined}}%
\hbox{\csname b@\@citeb\endcsname}}}{#1}}

\def\citer{\@ifnextchar [{\@tempswatrue\@citexr}{\@tempswafalse\@citexr[]}}

\def\@citexr[#1]#2{\if@filesw\immediate\write\@auxout{\string\citation{#2}}\fi
  \def\@citea{}\@cite{\@for\@citeb:=#2\do
    {\@citea\def\@citea{--\penalty\@m}\@ifundefined
       {b@\@citeb}{{\bf ?}\@warning
       {Citation `\@citeb' on page \thepage \space undefined}}%
\hbox{\csname b@\@citeb\endcsname}}}{#1}}
\catcode`@=12

\def\leftrightarrowfill{$\mathsurround=0pt \mathord\leftarrow \mkern-6mu
        \cleaders\hbox{$\mkern-2mu \mathord- \mkern-2mu$}\hfill
        \mkern-6mu \mathord\rightarrow$}       % <--> double differential
\def\dvec#1{\vbox{\ialign{##\crcr
        \leftrightarrowfill\crcr\noalign{\kern-1pt\nointerlineskip}
        $\hfil\displaystyle{#1}\hfil$\crcr}}}           % <--> accent
\def\beq{\begin{equation}}
\def\eeq{\end{equation}}

\def\beqx{\begin{displaymath}}
\def\eeqx{\end{displaymath}}

\def\beql{\begin{eqnarray}}
\def\eeql{\end{eqnarray}}

\def\Journal#1#2#3#4{{#1} {\bf #2} (#4) #3}

\def\NPB{{\em Nucl. Phys.} B}
\def\NPPS{\em Nucl. Phys. (Proc. Suppl.)} 
\def\PLB{{\em Phys. Lett.} B} 
\def\PRL{\em Phys. Rev. Lett.} 
\def\PRD{{\em Phys. Rev.} D} 
\def\ZPC{{\em Z. Phys.} C} 
\def\PR{\em Phys. Rev.}

\def\PR{\em Phys. Rep.}

\newcommand{\lwig}{\mbox{\,\raisebox{.3ex}
    {$<$}$\!\!\!\!\!$\raisebox{-.9ex}{$\sim$}\,}}
\newcommand{\gwig}{\mbox{\,\raisebox{.3ex}
    {$>$}$\!\!\!\!\!$\raisebox{-.9ex}{$\sim$}}\,}

\newcommand{\iai}{I\overline{I}}

\newcommand{\ii}{{\rm i}} 
 
\newcommand{\xpr}{{x^\prime}}

\begin{document}
\title{
{\normalsize\rightline{DESY 98-200}\rightline{hep-ph/9812359}} 
\vskip 1cm 
      \bf QCD-Instantons at HERA -- An Introduction\thanks{Plenary
      talk at the 3rd UK Phenomenology Workshop on HERA Physics, Sep 20-25,
      1998, Durham/UK.}
                          \\
  \vspace{11mm}}
\author{A. Ringwald and F. Schrempp\\[4mm] 
Deutsches Elektronen-Synchrotron DESY, Hamburg, Germany}
\begin{titlepage} 
  \maketitle
\begin{abstract}
We review our ongoing theoretical and phenomenological study of the
discovery potential for instanton-induced DIS events at HERA.
Constraints from recent lattice simulations will
be exploited and translated  into a ``fiducial'' kinematical region
for our  predictions of the instanton-induced DIS cross-section.
\end{abstract} 
\thispagestyle{empty}
\end{titlepage}
\newpage \setcounter{page}{2}

\section{Instantons}
Non-abelian gauge theories like QCD are known to exhibit a rich vacuum 
structure. The latter includes {\it topologically} non-trivial
fluctuations of the gauge fields, carrying an integer topological
charge $Q$.  The simplest building blocks  of topological structure
are  instantons ($Q=+1$) and anti-instantons  ($Q=-1$) which are well-known
explicit solutions of the euclidean field equations in four
dimensions~\cite{bpst}.   

Instantons ($I$) are widely believed to play an important r{\^o}le in various 
{\it long-distance} aspects~\cite{ssh} of QCD: 

First of all, they may provide a solution of the famous $U_A(1)$ 
problem~\cite{th} ($m_{\eta^\prime}\gg m_{\eta}$), with the corresponding 
pseudoscalar mass splitting related to the topological susceptibility 
in the pure gauge theory by the well-known Witten-Veneziano formula~\cite{wv}.
Moreover, a number of authors have attributed a strong
connection  of instantons with chiral symmetry breaking~\cite{dia,ssh}
as well as the hadron and glueball spectrum.

However, there are also very important {\it short-distance}
implications~\cite{bb,rs,mrs1,rs-pl,mrs2} of QCD instantons to which the
present report is devoted:

Instantons are known to induce certain processes  
which violate {\it chirality} in accord with the general
axial-anomaly relation~\cite{th} and which are forbidden in conventional
perturbation theory. Of particular interest in this context is the
{\it deep inelastic scattering} (DIS) regime. Here, 
hard in\-stan\-ton-induced processes may both be {\it
calculated}~\cite{mrs1,rs-pl,mrs2} within {\it instanton-per\-tur\-ba\-tion
theory}  and possibly be {\it detected
experimentally}~\cite{rs,grs,rs2,dis97-phen}. 
As a  key feature it has recently been shown~\cite{mrs1}, that in 
deep-inelastic scattering (DIS) the 
generic hard scale ${\cal Q}$ cuts off instantons with {\it large size} 
$\rho\gg {\cal Q}^{-1}$, over which one has no control theoretically. 

Our finalized results~\cite{rs-pl,mrs2} for inclusive instanton-induced DIS
cross-sections are summarized in sections~2 and 4. Their weak residual
renormalization-scale dependence is quite remarkable.

As a second main point of this review (section~3), constraints from
recent lattice simulations will be exploited~\cite{rs-pl,rs-pub}
and translated  into a ``fiducial'' kinematical region
for our  predictions of the instanton-induced DIS cross-section based
on  instanton-perturbation theory. In section~5 we discuss    
the expected event signature and search strategies based on our Monte 
Carlo generator~\cite{grs} QCDINS 1.60. Finally (section~6), we briefly address
an interesting class of ``fireball'' events, observed in photoproduction, 
in the context of instantons and put forward a promising 
proposal~\cite{rs-pub} on extending our theoretical predictions beyond
the regime of strict instanton perturbation theory.

\section{DIS cross-sections in instanton-perturbation theory}

In $I$-perturbation theory one expands the relevant Green's functions
about the known, classical instanton solution
$A_\mu=A^{(I)}_\mu+\ldots$ instead of  
the usual (trivial) field configuration $A^{(0)}_\mu=0$ and obtains a
corresponding set of modified Feynman rules. Like in conventional
pQCD, the gauge coupling $\alpha_s$ has to be small.  

\begin{figure}[ht]
\begin{center}
\epsfig{file=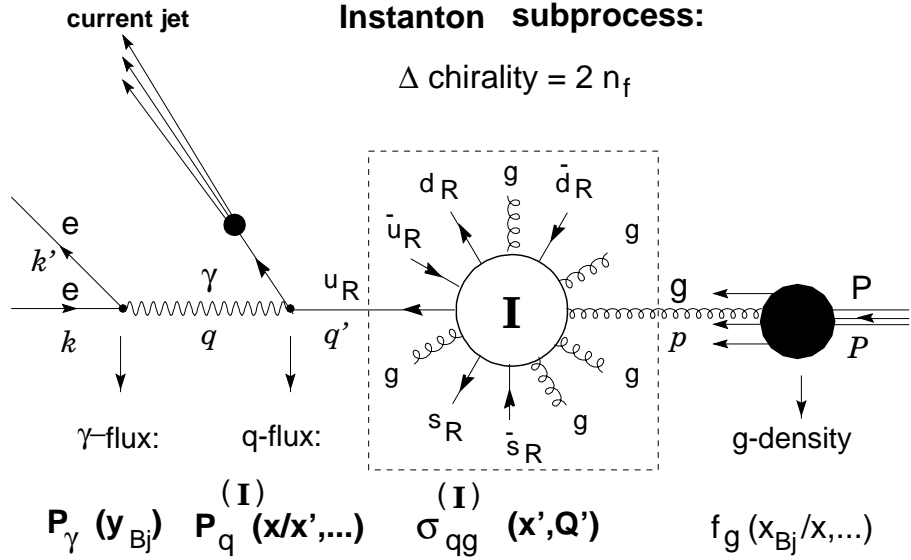,width=12cm}
\caption[dum]{\label{ev-displ}
The leading instanton-induced process in the DIS regime of $e^\pm$\,P 
scattering ($n_f=3$).}
\end{center}
\end{figure}

The leading instanton-induced process in the DIS regime of $e^\pm
P$ scattering for large photon virtuality $Q^2$ is illustrated in 
figure~\ref{ev-displ}. The dashed box emphasizes the  so-called  
%%%%%%%%%%%%%%%%%%%%%%%%%%%%%FIGURE  %%%%%%%%%%%%%%%%%%%%%%%%%%%%%%%%%
%%%%%%%%%%%%%%%%%%%%%%%%%%%%%%%%%%%%%%%%%%%%%%%%%%%%%%%%%%%%%%%%%%%%%%  
instanton-{\it subprocess} with its own Bjorken variables,
\begin{equation}
Q^{\prime\,2}= -q^{\prime\,2}\ge 0;\hspace{2ex}
\xpr=\frac{Q^{\prime\,2}}{2 p\cdot  q^\prime}\le 1.
\end{equation}
It induces a total chirality violation $\Delta\,{\rm chirality} = 2\,n_f$, in
accord with the corresponding axial anomaly~\cite{th}.
In the Bjorken limit of $I$-perturbation theory,
the dominant $I$-induced contribution to the inclusive HERA cross-section 
may be shown to  take the form~\cite{rs-pl,mrs2}
\begin{equation}
\frac{{\rm d}\sigma_{\rm  HERA}^{({I})}}{{\rm d}\xpr {\rm
d}Q^{\prime\,2}}\simeq \frac{{\rm d}{\cal L}_{q\,g}^{({
I})}}{{\rm d}\xpr {\rm d}Q^{\prime\,2}}\cdot 
\sigma_{q\,g}^{({I})}(Q^\prime,\xpr).
\label{ePcross}
\end{equation}

The differential luminosity, ${\rm d}{\cal L}_{q\,g}^{(I)}$, accounting
for the number of $q\,g$ collisions per $eP$ 
collision, has a convolution-like structure. It involves integrations
over  the gluon density, the $\gamma$-flux ${\rm
P}_{\gamma}$ and  the known $q$-flux ${\rm P}_{q}^{(I)}$
in the $I$-background (c.\,f. figure~\ref{ev-displ}). The crucial 
instanton-dynamics resides in the $I$-subprocess total cross-section 
$\sigma_{q\,g}^{(I)}(Q^\prime,\xpr)$, on which we focus our
attention  next~\cite{rs-pl,mrs2}.

Being an observable, $\sigma_{q\,g}^{(I)}(Q^\prime,\xpr)$
involves integrations over all $I (\overline{I})$-``collective
coordinates'',  including the $I\ (\overline{I})$-sizes 
$\rho\ (\overline{\rho})$ and the $\iai$-distance\footnote{Both an
instanton and an anti-instanton enter here, 
since cross sections result from taking the {\it modulus squared} of an
amplitude in the single $I$ background.} 4-vector $R_\mu$,
\begin{equation}
\sigma_{q\,g}^{(I)}=
      \int\limits_0^\infty d\rho\, D(\rho)
      \int\limits_0^\infty d\overline{\rho}\, D(\overline{\rho})
      \int d^4 R\, \{\ldots\}
      {\rm e}^{-Q^\prime(\rho+\overline{\rho})}\,
      {\rm e}^{\ii\, (p+q^\prime)\cdot R} 
      {\rm e}^{{-\frac{4\pi}{\alpha_s}}\,
      \Omega\left(\frac{R^2}{\rho\overline{\rho}},
      \frac{\overline{\rho}}{\rho} \right)}.
\label{cs}
\end{equation}
The $\rho (\overline{\rho})$-integrals in (\ref{cs}) involve as generic weight
the $I(\overline{I})$-density 
$D(\rho(\overline{\rho}))$~\cite{th,ber,morretal},  
\begin{eqnarray}
D({\rho})&=&
\frac{d}{\rho^5} \left(\frac{2\pi}{\alpha_s(\mu_r)}\right)^{2\,N_c}
\exp{\left(-\frac{2\pi}{\alpha_s(\mu_r)}\right)}(\rho\, \mu_r)^{\beta_0+ 
\frac{\alpha_s(\mu_r)}{4\pi}(\beta_1-4\,N_c\beta_0)}\label{dens}\\
 d&=&\frac{2\,{\rm e}^{5/6}}{\pi^2\,(N_c-1)!(N_c-2)!}\,{\rm
 e}^{-1.51137\,N_c+0.29175\,n_f}
 \ \ (\mbox{$\overline{\rm MS}$ scheme});\nonumber\\
 \beta_0&=&\frac{11}{3}N_c-\frac{2}{3}n_f;\ \beta_1=
\frac{34}{3}N_c^2-\left(\frac{13}{3}N_c-\frac{1}{N_c}\right)\,n_f,\nonumber
\end{eqnarray}
with renormalization scale $\mu_r$ and $N_c=3$.

The function  $\Omega(R^2/(\rho\overline{\rho}),\ldots)$ in equation
(\ref{cs}), appearing in the exponent with a large numerical
coefficient $4\pi/\alpha_s$, incorporates the effects of final-state
gluons. Within strict $I$-perturbation theory, it is given in form
of a perturbative expansion~\cite{holypert}, while in the so-called
$\iai$-valley approximation~\cite{yung,valley-most-attr-orient}
$\Omega$ is associated  with an analytically known 
closed expression~\cite{valley-most-attr-orient,valley-gen-orient} for
the interaction between $I$ and $\bar{I}$,       
$\Omega\simeq \alpha_s/(4\pi)S[A^{\iai}_\mu]-1$.
With both methods agreeing for larger values of
$R^2/(\rho\overline{\rho})$, we have actually used the valley method
in  our quantitative evaluation.  

Due to the nonvanishing virtuality $Q^{\prime\,2}$ in
DIS, the ``form factor'' $\exp{[-Q^\prime(\rho+\overline{\rho})]}$ in
(\ref{cs}), being associated with the off-shell quark (zero mode) $q$,
suppresses  large-size instantons~\cite{mrs1,rs-pl,mrs2}. Hence, the
integrals in (\ref{cs}) are {\it finite}.
In fact, they are dominated by a  unique {\it saddle-point}~\cite{rs-pl,mrs2},
\begin{equation}   
\rho^\ast = \overline{\rho}^\ast\sim 1/Q^\prime;\hspace{0.5cm} 
R^{\ast 2}\sim 1/(p+q^\prime)^2\ \Rightarrow \ 
\frac{R^\ast}{\rho^\ast}\sim \sqrt{\frac{\xpr}{1-\xpr}},
\label{saddle}
\end{equation}
from which it becomes apparent that the virtuality $Q^\prime$ controls the
effective $I$-size, while $\xpr$ determines the effective
$\iai$-distance (in units of the size $\rho$).  

\begin{figure}[b]
\begin{center}
\epsfig{file=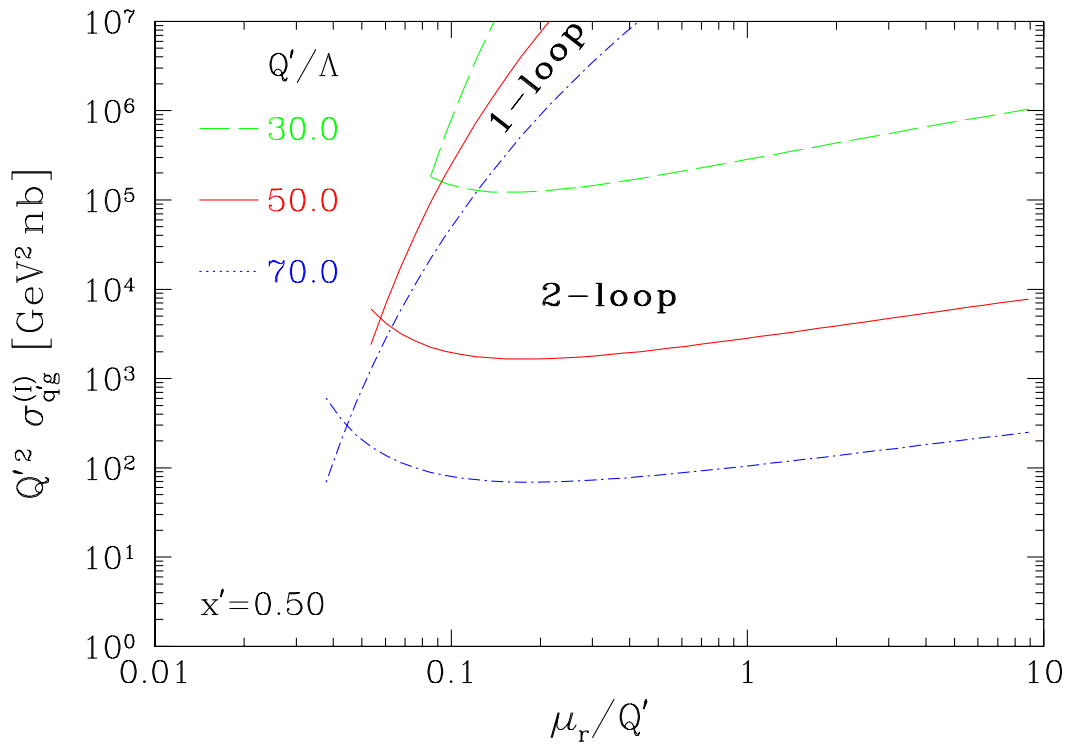,angle=270,width=14cm}
\caption[dum]{\label{renorm}
Illustration of the weak residual renormalization-scale ($\mu_r$)
    dependence of the resulting $I$-subprocess  cross-section
    $\sigma_{q\,g}^{(I)}(Q^\prime,\xpr)$.} 
\end{center}
\end{figure}

In figure~\ref{renorm}, the resulting 
$I$-subprocess cross-sections~(\ref{cs}) is displayed~\cite{rs-pl} over a {\it
large}  range of $\mu_r/Q^\prime$ for fixed $\xpr=0.5$ and
$Q^\prime/\Lambda=30,50,70$.
%%%%%%%%%%%%%%%%%%%%%%%%%%%%%FIGURE  %%%%%%%%%%%%%%%%%%%%%%%%%%%%%%%%%
%%%%%%%%%%%%%%%%%%%%%%%%%%%%%%%%%%%%%%%%%%%%%%%%%%%%%%%%%%%%%%%%%%%%%%  
Apparently, we have achieved great progress 
in stability and hence predictivity by using the improved  
expression~(\ref{dens})
of the $I$-density $D(\rho)$, which is renormalization-group (RG) invariant
at the {\it 2-loop } level, i.e. $D^{-1}\,{\rm d}D/{\rm
d}\ln(\mu_r)=\mathcal{O}(\alpha_s^2)$. The residual dependence on the
renormalization scale $\mu_r$ is remarkably flat and turns out to be
strongly reduced as compared to the 1-loop case! Throughout, we
choose as the ``best scale'', $\mu_r = 0.15\ Q^\prime$, for which 
$\partial \sigma^{(I)}_{q\,g}/\partial \mu_r \simeq 0$
(c.\,f. figure~\ref{renorm}). This choice agrees well with the
intuitive expectation~\cite{mrs1,bb} $\mu_r \sim 1/\langle \rho
\rangle \sim Q^\prime/\beta_0 ={\cal O}(0.1)\, Q^\prime$.  
%%%%%%%%%%%%%%%%%%%%%%%%%FIGURE%%%%%%%%%%%%%%%%%%%%%%%%%%%%%%%%%%%%%
\begin{figure}[b]
\begin{center}
\parbox{5.3cm}{\epsfig{file=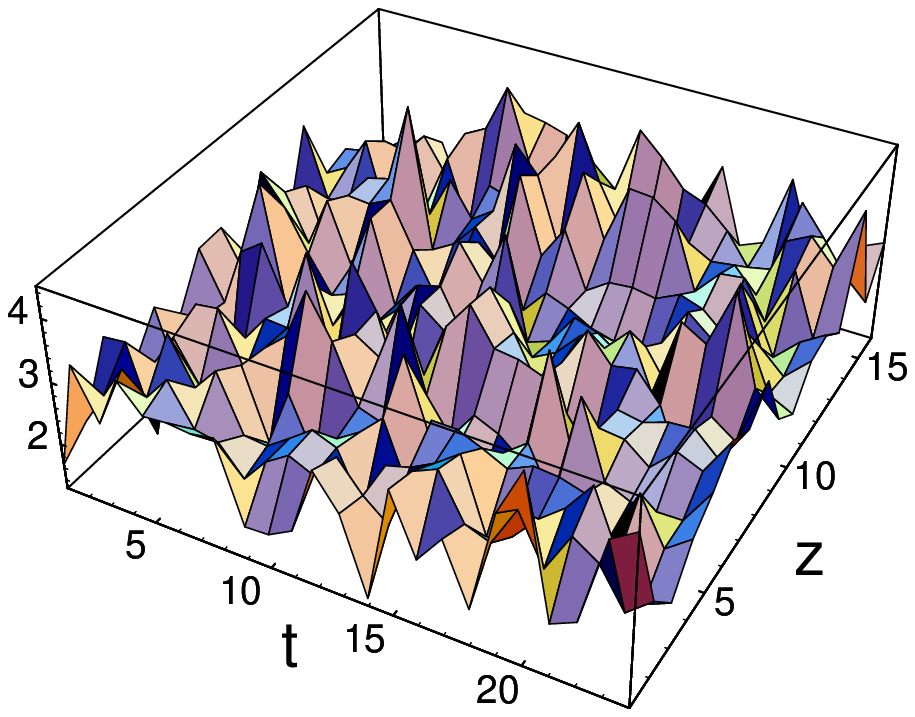,width=5.3cm}}
\parbox{5.725cm}{\epsfig{file=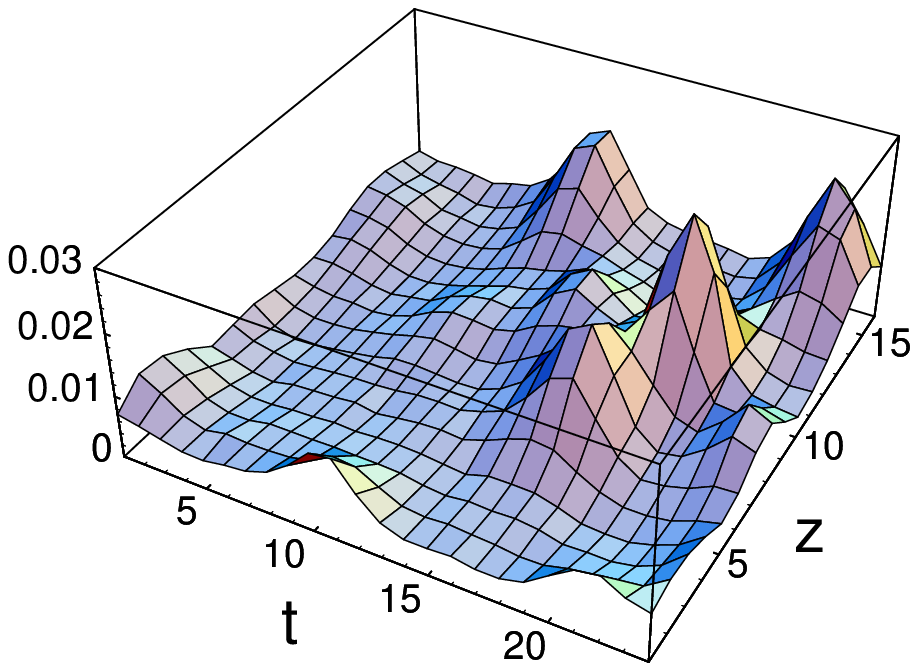,width=5.725cm}}
\parbox{5.725cm}{\epsfig{file=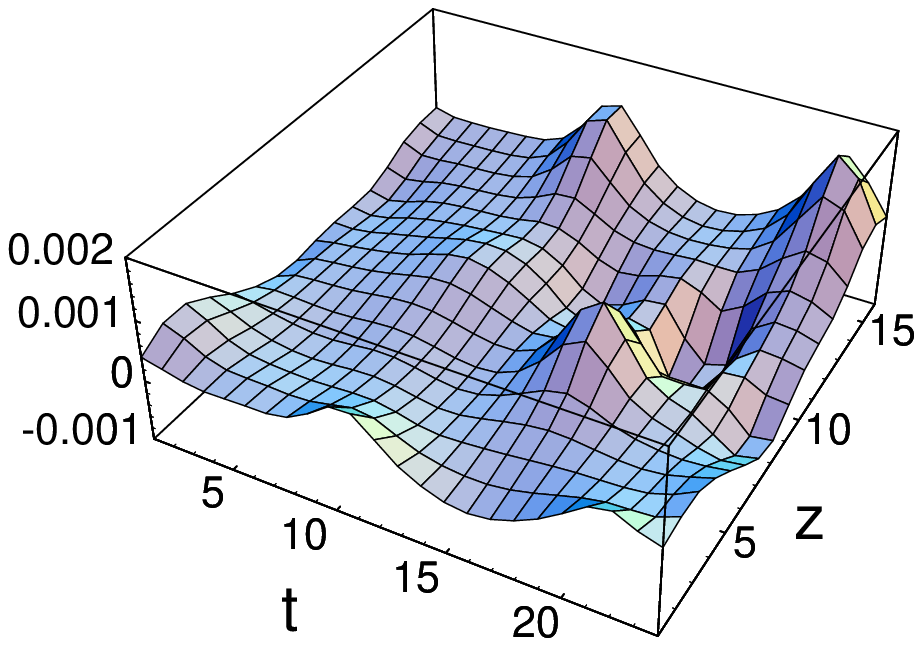,width=5.725cm}}
\put(-150,70){\small\bf Topological Charge Density}
\put(-370,70){\small\bf Lagrange Density}
\caption[dum]{\label{lattice-intro}Instanton content of a typical
slice of a gluon configuration at fixed x,y as a function of z and
t~\cite{chu}. (a) Lagrange 
density before ``cooling'', with  fluctuations of {\it short} wavelength ${\cal
O}(a)$ dominating. After ``cooling'' by 25 steps, 3 $I$'s and 2
$\overline{I}$'s may be clearly identified in the lagrange density (b) 
and the topological charge density (c).}
\end{center}
\end{figure}
%%%%%%%%%%%%%%%%%%%%%%%%%%%%%FIGURE%%%%%%%%%%%%%%%%%%%%%%%%%%%%%%%%%%%%%

\section{``Fiducial'' region from lattice simulations}

There has been much recent activity in the lattice 
community to ``measure'' topological fluctuations in 
lattice simulations~\cite{lattice} of QCD. Being
independent of perturbation theory, such simulations provide
``snapshots'' of the QCD vacuum including all possible
non-perturbative features like instantons (figure~\ref{lattice-intro}).  
Let us discuss next, how these lattice results may be exploited to
provide crucial support for the theoretical basis of our calculations
in DIS: 

To this end, we first perform a {\it quantitative}
confrontation~\cite{rs-pl,rs-pub} of the 
predictions from $I$-perturbation theory with a recent high-quality lattice
simulation~\cite{ukqcd} of QCD (without fermions, $n_f=0$). The
striking agreement which we shall find over a range of $I$-collective
coordinates is a very interesting result by itself. 

Next, we recall (c.\,f. (\ref{cs}) and (\ref{saddle})) that the
collective coordinate integrals in our DIS cross-section
$\sigma_{q\,g}^{(I)}(Q^\prime,\xpr)$  are dominated by a
unique, calculable saddle-point ($\rho^\ast,R^\ast/\rho^\ast$), in one-to-one
correspondence to the conjugate momentum variables ($Q^\prime ,\xpr$).  
This fact then allows us to {\it translate} the extracted range of validity
of $I$-perturbation theory and the dilute $I$-gas approximation, 
($\rho\leq \rho_{\rm max}, R/\rho\geq (R/\rho)_{\rm min}$), 
directly into a ``fiducial'' kinematical region  
($Q^\prime\geq Q^\prime_{\rm min},\xpr\geq x^\prime_{\rm min}$) in
momentum space!

In lattice simulations 4d-Euclidean space-time is made discrete;
specifically, the recent ``data''
from the UKQCD 
collaboration~\cite{ukqcd}, which we shall use here, involve a lattice
spacing $a = 0.055 - 0.1$ fm and a volume  $V=l_{\rm space}^{\,3}\cdot
l_{\rm time}=[16^3\cdot 48 - 32^3\cdot 64]\,a^4$.   
In principle, such a lattice allows to study the properties of an 
ensemble of $I$'s and $\overline{I}$'s  with sizes $a < \rho < V^{1/4}$. 
However, in order to make instanton effects visible, a certain ``cooling'' 
procedure has to be applied first. It is designed to
filter out (dominating) fluctuations of {\it short} wavelength ${\cal
O}(a)$ (c.\,f. figure~\ref{lattice-intro}~(a)), while affecting the
topological fluctuations  of much longer wavelength $\rho \gg a$
comparatively  little. After ``cooling'', $I$'s and $\overline{I}$'s
can clearly be seen (and studied) as bumps in the
lagrange density and the topological charge density
(figure~\ref{lattice-intro}~(b), (c)).  For a more detailed
discussion of  lattice-specific caveats, like possible lattice
artefacts and the dependence of results on ``cooling'' etc., see
Refs.~\cite{lattice,ukqcd}.    

\begin{figure} [b]
\begin{center}
\parbox{6.25cm}{\vspace{-0.3cm}\epsfig{file=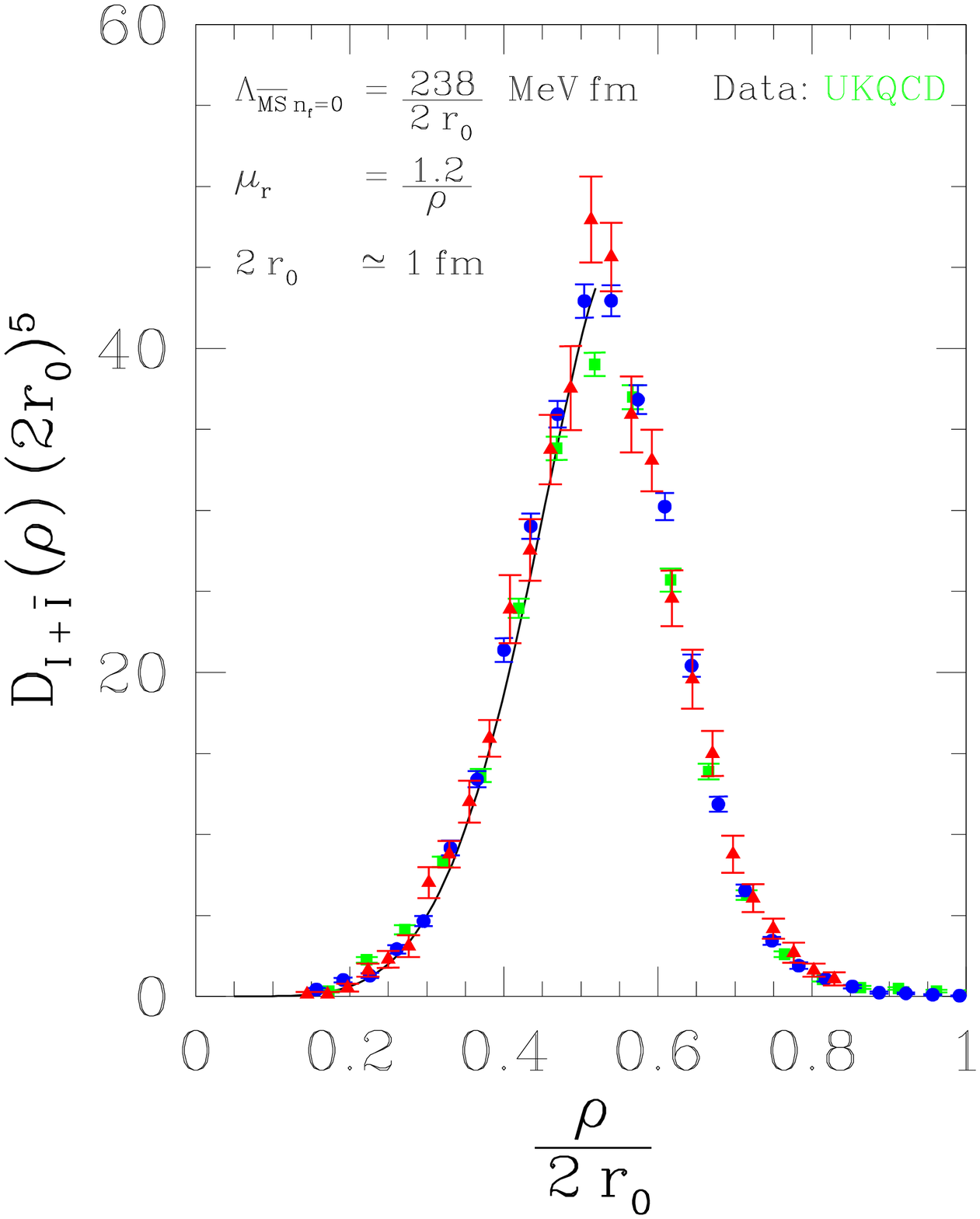,width=6.25cm}}
%\hspace{0.1cm}
\parbox{10.65cm}{\vspace{-0.3cm}\epsfig{file=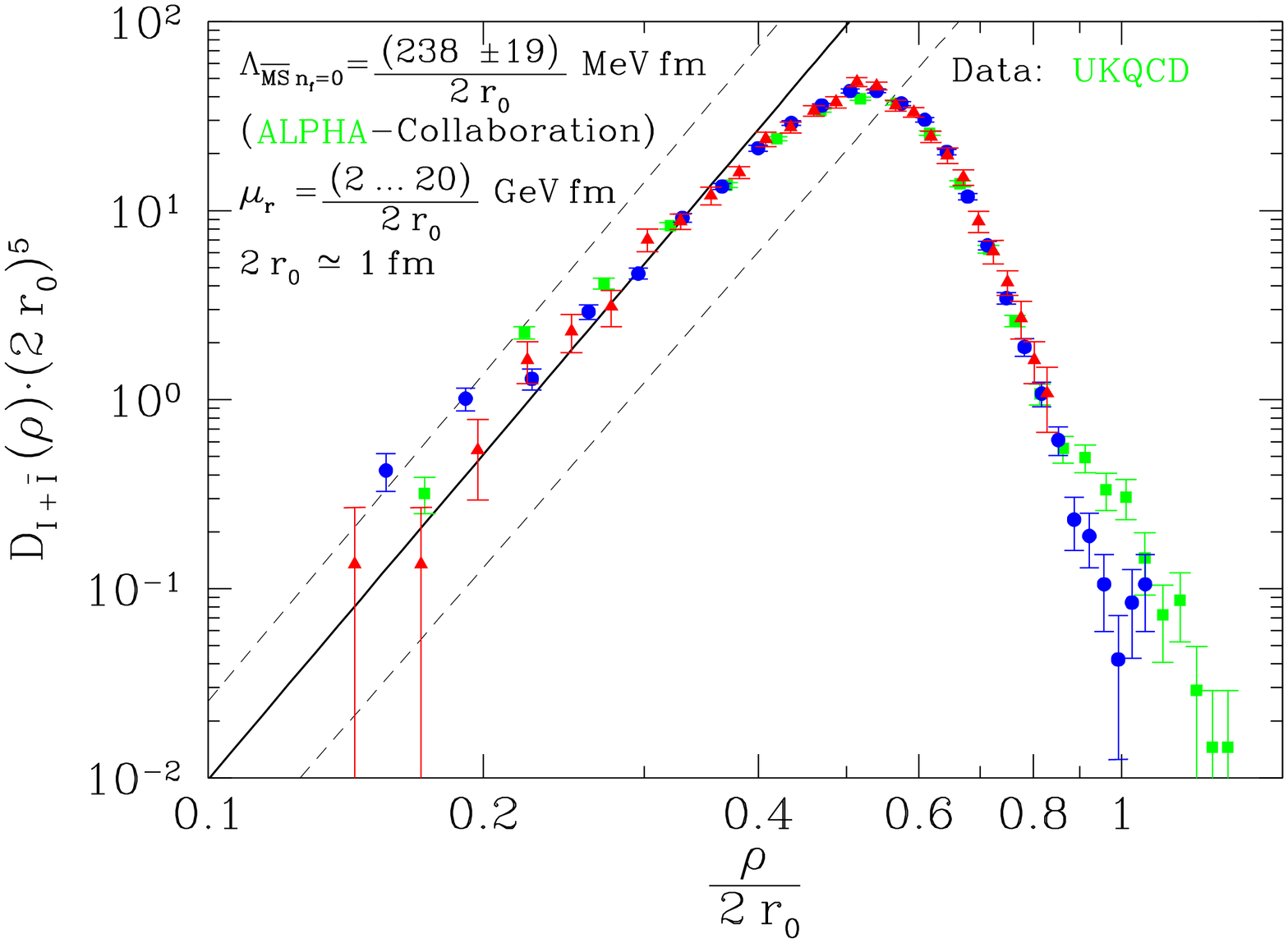,%
angle=-90,width=10.65cm}}
\end{center}
\caption[dum]{\label{lattice}
    Continuum limit~\cite{rs-pub} of ``equivalent'' UKQCD
    data~\cite{ukqcd} for the 
    ($I+\overline{I}$)-density at $(\beta,n_{\rm
    cools})=(6.0,23)\,[\Box ],\ (6.2,46)\,[\circ ],\
    (6.4,80)\,[\triangle ]$. The striking agreement with $2\,D(\rho)$      
    of $I$-perturbation theory from (\ref{dens}) is
    apparent~\cite{rs-pub}. The 3-loop form of 
    $\alpha_s$ with $\Lambda_{\overline{\rm MS}\,n_f=0}$ from
    ALPHA~\cite{alpha} was used.  
    (a) For $\mu_r=1.2/\rho$, the agreement extends up to the peak; 
    (b) Log-log plot to exhibit the expected power law $\sim \rho^6$
        and the agreement in magnitude for small $\rho$ over a wide
        range of $\mu_r$.  The dashed error band results from varying  
    $\Lambda_{\overline{\rm MS}\,n_f=0}$ and $\mu_r$ within its error and
    given range, respectively.}    
\end{figure}

Of course, one has to extrapolate the lattice observables to the
continuum ($a\Rightarrow 0$), before a meaningful comparison with
$I$-perturbation theory can be made. This is complicated by
a strong dependence of the various distributions on the number $n_{\rm cools}$ 
of cooling sweeps for {\it fixed} $\beta=6/g^2_{\rm lat}$. In
ref.~\cite{ukqcd}, however, {\it equivalent} pairs ($\beta,n_{\rm
cools}$) were found, for which {\it  shape} and {\it
normalization} of the distributions essentially remain {\it invariant}. For
instance, the continuum extrapolation of the data for the
($I+\overline{I}$)-density $D_{I+\overline{I}}$ at $(\beta,n_{\rm
cools})= (6.0,23),\ (6.2,46),\ (6.4,80)$,
may thus be performed quite reliably~\cite{rs-pub}, by simply
rescaling the arguments 
$\rho \Rightarrow\overline{\rho(0)}/\overline{\rho(a)}\cdot
\rho$. Here, $\overline{\rho(0)}$ denotes the continuum limit of the
{\it weakly varying} average $\rho$ values, $\overline{\rho(a)}$, of
$D_{I+\overline{I}}(\rho,a)$. A {\it linear}
extrapolation in $(a/r_0)^2$ was employed. For consistency and minimization of
uncertainties,  one should
use only a single dimensionful quantity to relate lattice units and
physical units. 
Throughout our analysis, all dimensions are therefore 
expressed by the so-called Sommer scale~\cite{sommer,alpha} $r_0$, with
$2\,r_0\simeq 1$ fm, which we prefer over the
string tension~\cite{ukqcd}. 
The resulting ``continuum data'' for
$D_{I+\overline{I}}(\rho)$ are displayed in figure~\ref{lattice}. They
scale nicely. 
%%%%%%%%%%%%%%%%%%%%%%%%%%%%%%%%%%%%%%%%%%%%%%%%%%%%%%%%%%%%%%%%%%%%
%%%%%%%%%%%%%%%%%%%%%%%%%%%%%%%%%%%%%%%%%%%%%%%%%%%%%%%%%%%%%%%%%%%%%%
We are now ready to perform a quantitative comparison with the predictions of
$I$-perturbation theory~\cite{rs-pub}. For reasons of space, let us
concentrate here on the ($I+\overline{I}$)-density
$D_{I+\overline{I}}(\rho)$. The prediction     
(\ref{dens}) of $I$-perturbation theory  is a power law
for {\it small} $\rho$, i.\,e. approximately $D \sim \rho^6$ for $n_f=0$. Due 
to its 2-loop RG-invariance the normalization of
$D_{I+\overline{I}}(\rho)$ is practically
independent of the renormalization scale $\mu_r$ over a wide range.
It is strongly and exclusively dependent on
$r_0\,\Lambda_{\overline{\rm MS}\,n_f=0}$, for which we take the most recent,
accurate result by the ALPHA-collaboration~\cite{alpha},
$ 2\,r_0\,\Lambda_{\overline{\rm MS}\,n_f=0}=(238\pm 19)$ MeV\,fm.
In figure~\ref{lattice}~(b) we display both this parameter-free
prediction from (\ref{dens}) of $I$-perturbation theory  and the continuum
limit of the  UKQCD data  in a log-log
plot, to clearly exhibit the expected power law in $\rho$. The
agreement in shape {\it and} normalization for $\rho\lwig
0.3\,(2\,r_0)\simeq 0.3$ fm is striking, indeed, 
notably in view of the often criticized ``cooling'' procedure and the 
strong sensitivity to  $\Lambda_{\overline{\rm MS}\,n_f=0}$.

By a similar analysis~\cite{rs-pub}, we were able to infer from the
``equivalent'' UKQCD lattice data  a range of validity $R/\rho \gwig
1$ of the valley expression for the $\iai$-interaction
$\Omega(R^2/(\rho\overline{\rho}),\ldots)$ in 
(\ref{cs}). Finally, we have confirmed~\cite{ukqcd,rs-pub} the
approximate validity of the 
dilute-gas picture for sufficiently {\it small}
instantons\footnote[2]{Note that the full ($I+\overline{I}$)-ensemble
without the size restriction is known {\it not} to be a dilute
gas~\cite{ukqcd,lattice}.} with $\rho
\lwig (0.3 - 0.5)$ fm. The latter results are based on the ``packing
fraction''~\cite{ukqcd} being $< 1$ and a test of 
the dilute-gas identity: $\langle Q^2\rangle  = N_{\rm tot}$. Here Q
is the topological charge and $N_{\rm tot}$ the total number 
of charges. 
These results strongly support the reliability of 
our calculations in DIS.

By means of the discussed 
saddle-point correspondence (\ref{saddle}), these lattice constraints
may be converted into a ``fiducial'' region for our 
cross-section predictions in DIS~\cite{rs-pl},
\begin{equation}
 \left.\begin{array}{lcccl}\rho^\ast&\leq& \rho^\ast_{\rm max}&\simeq& 
         0.3 {\rm\ fm};\\[1ex]
 \frac{R^\ast}{\rho^\ast}&\geq&\left(\frac{R^\ast}{\rho^\ast}\right)_{\rm min}
 &\simeq& 1\\
 \end{array}\right\}\Rightarrow
 \left\{\begin{array}{lclcl}Q^\prime&\geq &Q^\prime_{\rm min}&\simeq&
 8 {\rm\ GeV};\\[1ex]
 x^\prime&\geq &x^\prime_{\rm min}&\simeq &0.35.\\
 \end{array} \right .
\label{fiducial}
\end{equation}

\section{HERA cross-section}

\begin{figure} [h]
\begin{center}
\epsfig{file=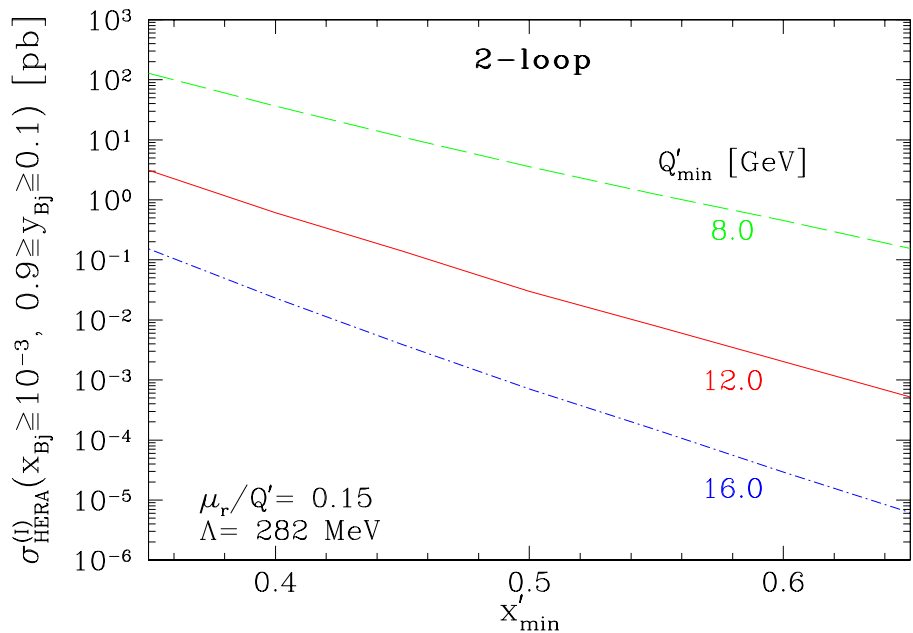,angle=270,width=12.0cm}
\caption[dum]{\label{cuts}
 $I$-induced cross-section at HERA as function of
 the cuts in ($\xpr,Q^\prime$).}
\end{center}
\end{figure}

Figure~\ref{cuts} displays our finalized $I$-induced cross-section at
HERA~\cite{rs-pl,mrs2},
as function of the cuts $x^\prime_{\rm min}$ and $Q^\prime_{\rm min}$.
%%%%%%%%%%%%%%%%%%%%%%%%%%%%%%%%%%%%%%%%%%%%%%%%%%%%%%%%%%%%%%%%%%%%
%%%%%%%%%%%%%%%%%%%%%%%%%%%%%%%%%%%%%%%%%%%%%%%%%%%%%%%%%%%%%%%%%%%%%%
For the minimal cuts (\ref{fiducial}) extracted from the UKQCD lattice
simulation, we obtain a surprisingly large cross-section, 
\begin{equation}
\label{minimal-cuts}
\sigma^{(I)}_{\rm HERA}(\xpr\ge0.35,Q^\prime\ge 8\, {\rm GeV}) 
\simeq 126\, {\rm pb};\ x_{\rm Bj}\ge 10^{-3};\ 0.9\ge y_{\rm Bj}\ge 0.1 .
\end{equation}
Hence, with the total luminosity accumulated by experiments at HERA, 
${\mathcal L}={\mathcal O}(80)$ pb$^{-1}$, one already expects  
${\mathcal O}(10^4)$ $I$-induced events on tape from this kinematical region. 
Note also that the cross-section quoted in Eq.~(\ref{minimal-cuts})
corresponds to a fraction of $I$-induced to normal DIS (nDIS) events of
\begin{equation}
f^{(I)} = \frac{\sigma^{(I)}_{\rm HERA}}{\sigma^{({\rm nDIS})}_{\rm HERA}}
={\mathcal O}(1)\, \%; 
\hspace{6ex} {\rm for}\ x_{\rm Bj}\ge 10^{-3};\ 0.9\ge y_{\rm Bj}\ge 0.1 .
\end{equation}
This is remarkably close to the published upper limits on the fraction of
$I$-induced events~\cite{limits}, which are also on the one percent
level.

There are still a number of significant uncertainties in our result
for the cross-section. 
For {\it fixed} $Q^\prime$ and $\xpr$ cuts, one of the dominant uncertainties
arises from the experimental uncertainty in the QCD scale $\Lambda$.  
In the 2-loop expression for $\alpha_s$ with $n_f=3$ (massless)
flavours we used the value $\Lambda_{\overline{\rm MS}}^{(3)}=282$ MeV, corresponding to the central value of the DIS average for $n_f=4$, 
$\Lambda_{\overline{\rm MS}}^{(4)}=234$ MeV~\cite{pdg}. If we change 
$\Lambda_{\overline{\rm MS}}^{(3)}$ within the allowed range, 
$\approx\pm 65$ MeV, the cross-section (\ref{minimal-cuts}) varies
between 26 pb and 426 pb. Minor uncertainties are associated with  
the residual renormalization-scale dependence
(c.f. figure~\ref{renorm}) and the choice of the factorization
scale. Upon varying the latter by an order of 
magnitude, the changes are in the ${\mathcal O}(20)$ \% range only.

By far the dominant uncertainty in $\sigma^{(I)}_{\rm HERA}$ arises,
however, from the uncertainty in placing the ($\xpr ,Q^\prime$) cuts  
(c.f. figure~\ref{cuts}). Hence, the constraints (\ref{fiducial}) from 
lattice simulations are extremely valuable for making concrete and
reliable predictions of the $I$-induced rate at HERA.

\section{Signatures and searches}

An indispensable tool for investigating the structure of the
$I$-induced final state and for developping optimized search
strategies is our Monte-Carlo generator for $I$-induced DIS-events,
QCDINS 1.60. Besides the matrix element for the $I$-induced hard
subprocess, it provides leading-log parton showers and hadronization
via its interface to HERWIG 5.9. 

The characteristic features of the $I$-induced final state are
illustrated in figure~\ref{events}~(a) displaying the lego plot of a
typical event from QCDINS 1.60 (c.\,f. also figure~\ref{ev-displ}):

%%%%%%%%%%%%%%%%%%%%%%%%%%%%%FIGURE%%%%%%%%%%%%%%
\begin{figure}[ht]
\vspace{-0.1cm}
\begin{center}
\parbox{7.5cm}{\epsfig{file=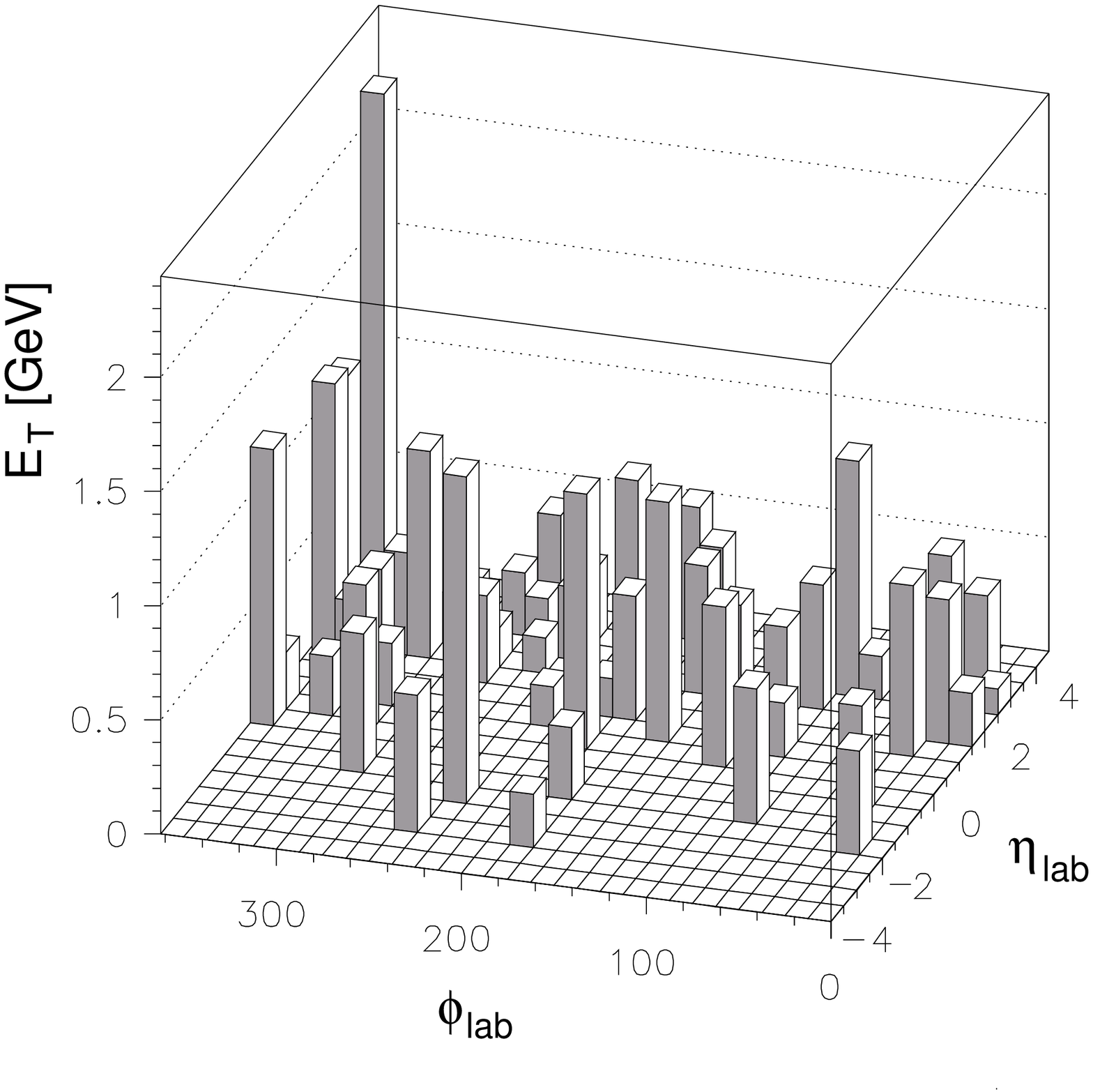,width=7.5cm}}
\hspace{0.7cm}
\parbox{7.5cm}{\vspace{-0.3cm}\epsfig{file=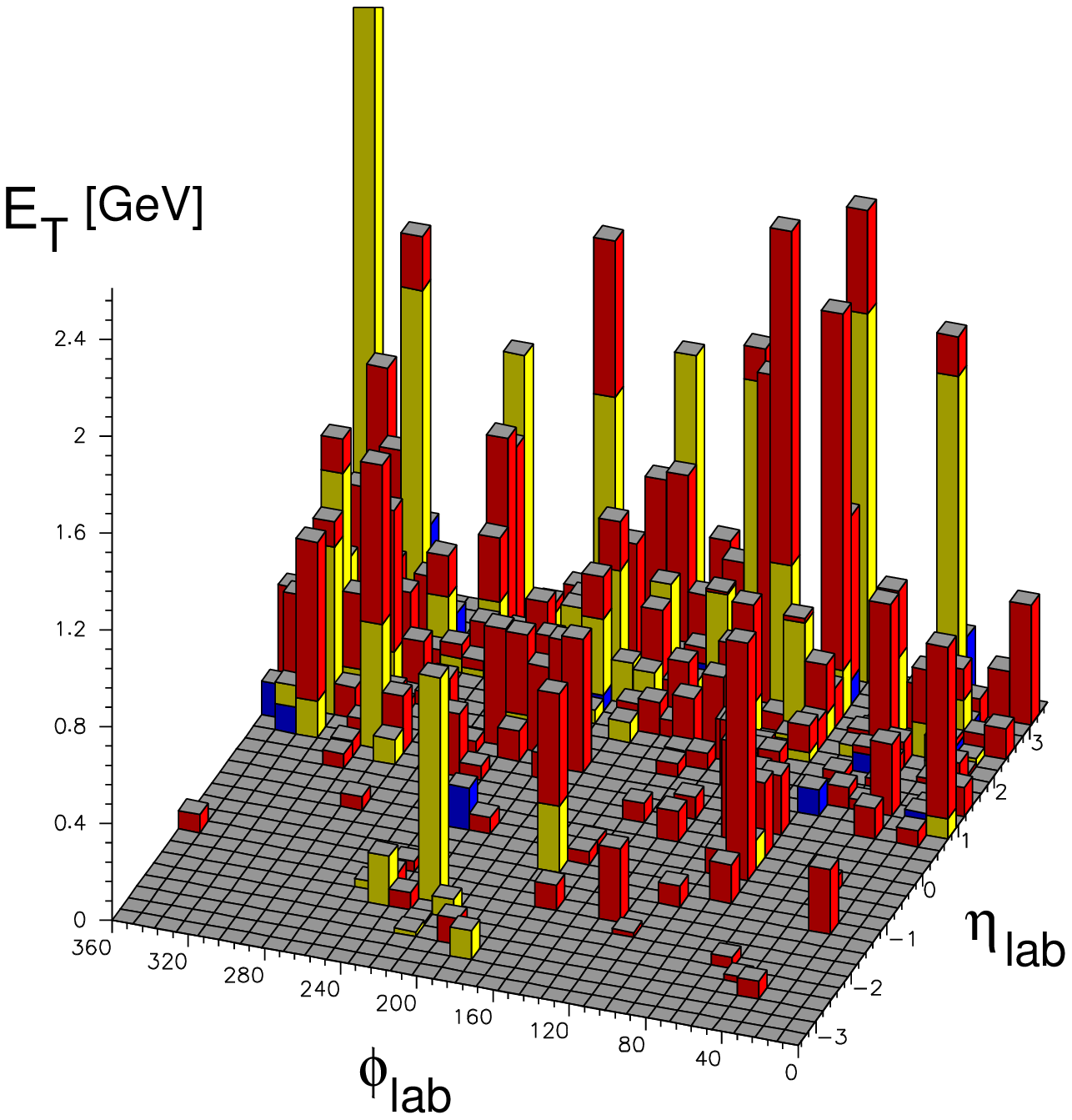,width=7.5cm}}
\caption[dum]{\label{events}
(a) Lego plot of a typical instanton-induced event from QCDINS 1.60.\\
(b) An interesting {\it real} ``fireball'' event in photoproduction 
    from ZEUS~\cite{fireball} with very large total $E_T$ and multiplicity.}
\end{center}
\end{figure}
%%%%%%%%%%%%%%%%%%%%%%%%%%%%%%%%%%%%%%%%%%%%%%%%%
Besides a single (not very hard) current-quark jet, one expects  an
accompanying densely populated {\it ``hadronic band''}. 
For $x_{\rm Bj\,min}\simeq
10^{-3}$, say,  it is centered around $ \overline{\eta} \simeq 2$  and
has a width of $\Delta \eta \simeq \pm 1$. The band directly reflects the {\it
isotropic}  production of an $I$-induced ``fireball'' of ${\cal
O}(10)$  partons in the $I$-rest system. Both the total transverse
energy $\langle E_T \rangle \simeq 15$ GeV and the charged particle
multiplicity $\langle n_c\rangle \simeq 13$ in the band are far higher than in
normal DIS events. Finally, each $I$-induced event has to contain
strangeness (and possibly also charm)  such that the number of $K^0$'s amounts to $\simeq 2.2$/event. 

Despite the high expected rate (\ref{minimal-cuts}) of $I$-induced
events at HERA,  no {\it single} observable is known (yet) with
sufficient nDIS rejection. Hence, a dedicated multi-observable
analysis seems to be required. Neural network filters are being tried
and exhibit a very good analyzing power if applied to $\gwig {\cal
O}(5)$ observables~\cite{jgerigk}. Strategies to produce
``instanton-enriched'' data samples and to reconstruct
($Q^{\prime\,2},\xpr$) are under study and look quite
promising~\cite{jgerigk}.  Clearly, in all cases, a good
understanding of the perturbative QCD background in the {\it tails} of
the considered distributions is required.  

\section{Going beyond instanton-perturbation theory}

A class of striking ``fireball'' events in
photoproduction, with large
total $E_T$ and large multiplicity, has been reported~\cite{wg2wg3}
at this meeting (see e.\,g. figure~\ref{events}~(b)). While the
quantitative analysis is still in an early stage, these events seem
to exhibit all characteristics of $I$-induced events
(c.\,f. figure~\ref{events}~(a)).  
However, it appears that -- unlike ordinary QCD perturbation theory -- the
hard photoproduction limit, $Q^2 \Rightarrow 0,\ (E_T)_{\rm jet}$
large, is not within the reach of strict $I$-perturbation theory. 
The reason is that in the $Q^2 \Rightarrow 0$ limit, we encounter a 
contribution to the photoproduction cross-section, which tends to {\it
diverge}, if integrated  over the $I$-size $\rho$. This IR divergence
at large $\rho$ is independent of the $E_T$ of the (current quark) jet and
directly associated with the ``bad'' large-$\rho$ behaviour of the
{\it perturbative} expression (\ref{dens}) for the $I$-density, 
$D(\rho) \sim \rho^{6-2/3 n_f}$. In contrast, the {\it actual} form of
$D(\rho)$ (c.\,f. figure~\ref{lattice}) is strongly peaked around
$\rho \simeq 0.5$ fm, and appears to {\it vanish} exponentially fast
for larger $\rho$.    
The above ``fireball'' events and more generally, the ongoing $I$-searches
at HERA, provide  plenty of motivation for trying to extend our
calculational framework beyond strict $I$-perturbation
theory. We are thus led to make the following promising proposal in
this direction: 

One may try and replace the most strongly varying entries in the
perturbative calculations, the $I$-density $D(\rho)$ and the
$\iai$-interaction $\Omega(R^2/(\rho\overline{\rho}),\ldots)$, in 
(\ref{cs}), by their {\it actual} form as extracted from the
recent non-perturbative lattice results~\cite{rs-pub}. 

The $I$-rates in photoproduction and the $I$-contributions to 
further interesting observables may then be calculated, and
the ($Q^{\prime\, 2},\xpr$) cuts in $I$-searches be considerably
relaxed! Due to the strong peaking of $D(\rho)$, only the region around
$\rho \simeq 0.5$ fm enters and  the dilute-gas
approximation may well continue to hold up to the peak (c.\,f. section~3).

\end{document}